\documentstyle[aps]{revtex}
\def\ba{\begin{eqnarray}}
\def\ea{\end{eqnarray}}

\begin{document}

\title{Fredholm methods for billiard eigenfunctions in the coherent state
representation}

\author{Fernando P. Simonotti and Marcos Saraceno}

\address{Departamento de F\'{\i}sica, Comisi\'on Nacional de Energ\'{\i}a
At\'omica, Av.\ Libertador 8250, 1429 Buenos Aires, Argentina.}

\maketitle

\begin{abstract}

We obtain a semiclassical expression for the projector 
onto eigenfunctions by means of the Fredholm theory. We express the projector 
in the coherent state basis, thus obtaining the semiclassical Husimi 
representation of the stadium eigenfunctions, which is written in terms of 
classical invariants: periodic points, their monodromy matrices and Maslov 
indices.


\end{abstract}

\section{Introduction}
\label{secintrofred}

The precise test of semiclassical approximations in the presence of 
chaos is of grest interest to establish the limits of applicability 
of periodic orbit theory and its resummations. This test can be done 
in model systems both on approximations to the spectrum or to the 
stationary states. For the calculation of the spectrum the most efficient
tool in this respect seems to be the spectral determinant 
\cite{bogo,vorosspec} and several calculations \cite{keating} 
have demonstrated that, given enough periodic orbits, the spectrum can 
be accurately represented semiclassically. However, a more sensitive 
test - and still a great challenge - is the semiclassical representation
of single eigenfunctions. This includes the study of the scar phenomena 
\cite{heller,bogoscars,berryscars,agam,smilansky,simonotti,kaplan} 
and  the 
eventual deviations from uniformity of eigenfunctions in accordance 
to the Berry-Voros hypothesis \cite{berryconj,vorosconj} and 
Schnirelman's theorem \cite{shnirelman}.

Just as for spectral problems, the use of Fredholm methods allows for 
the most efficient encoding of classical information in the calculation 
for single eigenfunctions \cite{agam,fishman}. 
In this paper we review these methods and apply
them to the calculation of Husimi distributions of stadium eigenfunctions.

This paper is organized as follows.
In Sec. \ref{secform} we review the Fredholm method for billiard 
eigenfunctions. Fredholm theory  allows us to find the solution to 
certain type of integral or operator equations \cite{smithies}.
For billiards these methods can be applied to the boundary integral equation.
In Sec. \ref{secapro} we make the semiclassical approximation that is based 
on the approximation of the traces and powers of the propagator as sums 
over the periodic points of the underlying classical system. The propagator
itself is taken as Bogomolny's ${\bf T}$ operator \cite{bogo}. 
We choose the coherent state
representation and obtain an expression for the semiclassical Husimi 
representation of the eigenfunctions in terms of classical invariants: 
periodic points, their monodromy matrices and Maslov indeces.
In Sec. \ref{secauto} we apply this scheme for the stadium billiard.
Our conclusions and perspectives are presented in Sec. \ref{secconclu}.

\section{Fredholm formulae for eigenfunctions}
\label{secform}

Fredholm theory gives the solution to a certain class of integral equations, 
which can also be written as operator equations \cite{smithies}. 
A Fredholm integral equation of second type is 
\ba
\chi(q)=\chi_{0}(q)+\lambda \int dq' {\bf T}(q',q) \chi(q').
\ea 
All the functions are defined in a finite domain. If the known functions 
$\chi_{0}(q)$ and ${\bf T}(q',q)$ are well behaved, the Fredholm alternative 
holds: there is a unique solution $\chi$ with the same analytic properties 
or the homogeneous equation ($\chi_{0}=0$) has a solution. There is a set 
of complex parameters $\lambda_i$ for which the solution is not unique. 
In operator notation, the inverse of $(1-\lambda {\bf T})$ exists if 
$\lambda \neq \lambda_{i}$. In this case, this inverse can be written as
\ba
\label{fredexp}
{1\over{1-\lambda {\bf T}}}={{{\bf M}(\lambda)}\over{D(\lambda)}},
\ea
where the operator ${\bf M}(\lambda)$ and the function $D(\lambda)$ are 
series in $\lambda$. If ${\bf T}$ is a compact operator, $D(\lambda)$ and
${\bf M}(\lambda)$ are entire in $\lambda$ and, thus, absolutely convergent.
The explicit form for the series expansion in terms of powers of ${\bf T}$ 
is given below. In what follows we apply this general theory 
assuming $\bf T$ to be unitary and of finite dimension $N$. Both 
assumptions are justified in the semiclassical limit for the 
quantization of billiards \cite{bogo}.

\subsection{Secular equation}

The $k$'s eigenvalues are given by the secular equation
$P(k)=\mbox{det}(1- {\bf T}(k))=0$. We can expand this determinant as
\ba
P(k)~=~\sum_{n=0}^{N} \beta_{n}(k),
\ea
where the coefficients $\beta_{n}(k)$ are related to the traces of 
${\bf T}(k)$, $b_{n}(k) \equiv \mbox{tr}  {\bf T}^{n}(k)$, through
\ba
\label{betac}
\beta_{n}(k)~=~-{1 \over {n}} \sum_{j=1}^{n} \beta_{n-j}(k) b_{j}(k).
\ea
Thus, knowledge of the traces up to a certain $n_{max}$ implies the 
knowledge of the coefficients $\beta_{n}$ up to the same $n_{max}$.

If ${\bf T}(k)$ is unitary, $P(k)$ is self 
reversive, meaning that its coefficients satisfy
\ba
\beta_{N-j}(k)~=~(-1)^{N}~ {\bar \beta}_{j}(k)~ \mbox{det} {\bf T}(k).
\ea

This condition alone forces the eigenvalues of $\bf T$ at fixed $k$ to be 
symmetric with respect to the unit circle: if $\lambda$ is an eigenvalue, 
then $1/{\bar \lambda}$ is an eigenvalue too. Of course, if $\bf T$ is 
unitary, then this condition is automatically satisfied but we can use it 
in our semiclassical approach to partially restore unitarity.

The contributions from coefficients $\beta_i$ with $i>[(N+1)/2]$ can be 
expressed in terms of coefficients
$\beta_i$ with $i \leq [(N+1)/2]$. ($[x]$ is the integer part of $x$.) 
So, if $N$ is even:
\ba
P(k)=\eta(k)+\det {\bf T} ~{\bar \eta}(k), \nonumber \\
\eta(k)=\sum_{j=0}^{N/2-1} \beta_{j}(k)+{1\over 2} \beta_{N/2}, 
~~
\det {\bf T}= {{\beta_{N/2}}\over{\bar \beta_{N/2}}}.
\ea 
If $N$ is odd:
\ba
P(k)=\eta(k)-\det {\bf T} ~{\bar \eta}(k), \nonumber \\
\eta(k)=\sum_{j=0}^{(N-1)/2} \beta_{j}(k), 
~~
\det {\bf T}= -{{\beta_{(N+1)/2}}\over{\bar \beta_{(N-1)/2}}}.
\ea 
As a consequence of the imposition of this symmetry on the operator 
${\bf T}$ we obtain two advantages: only traces up to half the 
Heisenberg time $t_{H}=N$ are needed and the eigenvalues are constrained
to lie on the unit circle or in symmetric pairs. 

These formulae relate $P(k)$ with the traces of powers of ${\bf T}(k)$
which, in turn are related semiclassically to periodic orbits and to
the smoothed density of states
\cite{tabor}. They have been tested extensively for the hyperbola 
billiard by Keating and Sieber \cite{keating}.  

\subsection{Green function}

To extend these methods to the calculation of eigenfunctions,
we define a generalized Green function ${\bf G}(k)$:
\ba
 {\bf G}(k)~=~{{{\bf T}(k)}\over{1~-~{\bf T}(k)}}.
\ea
This operator has poles at the billiard eigenvalues $k=k_{\nu}$ and its
residues are the projectors onto the corresponding eigenfunctions. 
It has a Fredholm expression as
\ba
{\bf G}(k)=
{{{\bf T}(k) {\bf C}^{t}({1- {\bf T}}(k))  }\over{P(k)}},
\ea
where $\bf{C}^{t}({1- {\bf T}}(k))$ is the transpose of the cofactor 
matrix of ${1- {\bf T}}(k)$ and, as in Eq. (\ref{fredexp}), has an 
expansion in powers of ${\bf T}(k)$.

It is then convenient to define a {\it normalized Green operator} as
\ba
{\bf g}(k)~=~{{{\bf G}(k)} \over {\mbox{tr}({\bf G}(k))}},
\ea
where the singularities in the denominator have been eliminated.
The normalized Green operator has the property ${\bf g}(k_{\nu})~=~
\left| \psi_{\nu} \rangle \langle \psi_{\nu} \right|$,
where $| \psi_{\nu} \rangle$ is the eigenvector corresponding to
eigenvalue $k_{\nu}$.
Then, we can write ${\bf g}(k)$ in the following way:
\ba
{ {\bf g}}(k)~=~{{ { {\bf T}}(k) ~C^{t}({1- {\bf T}}(k))
}\over{
\mbox{tr} ({ {\bf T}}(k) ~C^{t}({1- {\bf T}}(k))).
}}
\ea   

As the cofactor matrix can be expanded in powers of the propagator and as 
the propagator itself is unitary we write the normalized Green operator 
in terms of the powers of the propagator and their traces up to $N/2$ 
(if $N$ is even):
\ba
\label{proyec}
{ {\bf g}}(k)={{
\sum_{i=0}^{{N \over 2}-1} c_{i}(k) { {\bf T}}^{i+1}(k) -
\mbox{det}{ {\bf T}}(k)~
\sum_{i=0}^{{N \over 2}-1} {\bar c}_{i}(k) 
{ {\bf T}}^{\dagger~i}(k)   
} \over {
\sum_{i=0}^{{N \over 2}-1} c_{i}(k) \mbox{tr}({ {\bf T}}^{i+1}(k)) -
\mbox{det}{ {\bf T}}(k) ~
\sum_{i=0}^{{N \over 2}-1} {\bar c}_{i}(k) 
\mbox{tr}({ {\bf T}}^{\dagger~i}(k))
}} ,
\ea
where the coefficients $c_{i}(k)$ are given by
\ba
\label{cico}
c_{i}(k)~=~\sum_{n=i}^{{N \over 2}-1} \beta_{n-i}(k).
\ea
An analogous formula can be derived in case $N$ is odd.

The coefficients  $c_{i}(k)$ are dependent on the traces of ${\bf T}^{n}$
through Eqs. (\ref{betac}) and (\ref{cico}) and thus are
independent of the chosen representation.
On the other hand, the expression for the powers of the propagator will
depend on the representation chosen for the calculation of the 
eigenfunctions. If the coordinate representation $|q\rangle$ is chosen, 
Eq. (\ref{proyec}) relates the probability density $|\phi_{\nu}(q)|^2$ 
to the 
diagonal powers of the propagator $\langle q|{\bf T}^{n}(k)|q\rangle$.
If the Weyl representation is chosen, then Eq. (\ref{proyec}) gives the 
Wigner distribution of $|\phi_{\nu}\rangle$ in terms of the Weyl 
propagator \cite{agam,ozorio}. Here we choose the coherent state 
representation to find the equivalent Husimi distribution. 

We remark that Eq. (\ref{proyec}) is a very compact and representation
independent derivation of formulae that were previously very laboriously
derived for the Wigner case. It prepares in an optimal way the grounds 
for the semiclassical approximation because its ingredients are all 
dependent on classical elements, namely periodic orbits, phase space
volume and generating function.

Using the fact that at $k=k_{\nu}$ the normalized Green function is the 
projector onto the corresponding eigenstate, we can obtain the Husimi 
distribution as
\ba
\label{husimm}
{\cal H}_{\psi_{\nu}}(z,{\bar z})~=~
{{\langle z|{ {\bf g}}(k_\nu)|z \rangle} \over
{\langle z|z \rangle}}~=~{1\over{\langle z|z \rangle}} 
~{{
\sum_{i=0}^{{N \over 2}-1} c_{i}(k) 
\langle z| { {\bf T}}^{i+1}(k)|z \rangle  ~-~
\mbox{det}{ {\bf T}}(k)~
\sum_{i=0}^{{N \over 2}-1} {\bar c}_{i}(k) 
\langle z|{ {\bf T}}^{\dagger~i}(k)|z \rangle   
} \over {
\sum_{i=0}^{{N \over 2}-1} c_{i}(k) \mbox{tr}({ {\bf T}}^{i+1}(k)) ~-~
\mbox{det}{ {\bf T}}(k) ~
\sum_{i=0}^{{N \over 2}-1} {\bar c}_{i}(k) 
\mbox{tr}({ {\bf T}}^{\dagger~i}(k))
}}.
\ea

This scheme was successfuly applied in simple quantum maps
\cite{monastra}.

\section{Semiclassical approximation}
\label{secapro}

Green's theorem allows us to reduce the Schr\"odinger equation for the
billiard with Dirichlet boundary conditions to the following linear
homogeneous equation for the normal derivative on the border 
$\phi(s)$ 
\ba
\label{intecphi}
\phi(s) = -2 \oint ds' \phi(s') {\bf K}(s,s';k), 
\ea
where the kernel is
\ba
{\bf K}(s,s';k)
={{-ik}\over 2} \cos \psi(s)~ H_{1}^{(1)}(k|{\bf r}(s)-{\bf r'}(s')|),
\ea
with $k$ the wave number, $\psi(s)$ the angle between the normal at $s$ and 
the line that connects ${\bf r}(s)$ with ${\bf r'}(s')$ 
(see Fig. \ref{figbillar}) and $H_{1}^{(1)}$ 
the Hankel function of first type and order one.

We introduce the wave function $\mu(s)$
\ba
\phi(p)={1 \over {ik}}\sqrt{1-p^{2}} \mu(p),
\ea
with $\phi(p)$ and $\mu(p)$ the momentum representations of 
$\phi(s)$ and $\mu(s)$. 
This transformation makes the kernel symmetric and turns
Eq. (\ref{intecphi})  to
\ba
\label{intecbogo}
\mu(s)~=~\oint {\bf T}(s',s;k) \mu(s') ds'.
\ea

The semiclassical theory of the kernel ${\bf T}(s',s;k)$ \cite{bogo}
is based on two fundamental properties: ${\bf T}$ is semiclassically 
unitary and has an effective dimension $N(k)=Lk/\pi$, where $L$ is 
the length of the billiard. Moreover, the kernel is given by the 
generating function of the classical Birkhoff map (see Eq.(\ref{bogot})).
These properties have been extensively tested \cite{boasman} and will 
be assumed in what follows.

Thus, we make the semiclassical approximation by taking ${\bf T}$ 
as Bogomolny's 
operator (\ref{bogot}) and by evaluating all integrals by 
stationary phase approximation. The ${\bf T}$ operator for convex 
billiards in the plane, taking 
its border as Poincar\'e section and using Birkhoff coordinates is
\ba
\label{bogot}
{\bf T}(s',s;k)= 
\left({k \over {2 \pi i}}\right)^{ 1 \over 2}  
 \left|   {{\partial^2 l(s',s)} \over
{\partial s \partial s'}}  \right|^{1/2} \exp \left( i k l(s',s)
- i {\pi \over 2} \nu \right),
\ea
where the bounce map generated by $l(s's)$, the arc length between $s$
and $s'$; $\nu$ is the Maslov index. 
The quantization condition is $\mbox{det}(1-{\bf T}(k))=0$.

In the semiclassical theory for the spectral determinant
\ba
P(k)=\mbox{det}(1-{\bf T}(k))
\ea
the approximate unitarity of $\bf T$ can be used efficiently to reduce the 
number of periodic orbits needed for the computation of the spectrum.
Similar manipulations of the Fredholm formulae allow for the same reduction
in the semiclassical calculation of single eigenfunctions of the billiard.
We write down a formula giving the projector on  single eigenfucntions as a 
finite sum involving powers of the map $\bf T$ and its traces.

\subsection{Semiclassical traces and determinant}

First we need the traces. It is a well known fact that they adopt 
the following semiclassical expression \cite{tabor}:
\ba
[b_{n}]_{scl}=
\sum_{PO,n=n_{p}r}~
{{n_{p}} \over {|\mbox{det}(I- M_{p}^{r})|^{(1/2)} }}~
\exp \left( i r ( k l_{p} - \nu_{p} \pi/2 ) \right),
\ea
where the sum goes over all the primitive PO's of the   
billiard with period $n_{p}$, which must be a divisor of $n$, Maslov index 
$\nu_p$, length  $l_p$ and monodromy matrix $M_p$. The Maslov index can 
be interpreted geometrically: $\pi\nu_p$ is the angle swept by the 
unstable manifold of $M_p$ along the PO. 

The determinant of ${\bf T}$ can be obtained as \cite{bogo}
\ba
[\mbox{det}{ {\bf T}}(k)]_{scl}=(-1)^{N}~
\mbox{exp}(2 \pi i{\cal N}(k)),
\ea
where ${\cal N}(k)$, the number of states between 0 y $k$, is 
\ba
{\cal N}(k)={1 \over {4\pi}}{\cal A} k^2-{1 \over {4\pi}} L k,
\ea
with $\cal A$ the area of billiard and $L$ its length. 

These are the semiclassical ingredients needed for the calculation of the 
spectrum and of the coefficients $c_{i}(k)$.

\subsection{Semiclassical propagator in coherent state representation}

We first obtain the coherent state representation for one iteration of 
the bounce map:
\ba
\langle z'|{\bf T}|z\rangle=
\int~ds~ds'~\langle z'|s' \rangle 
~\langle s'|{\bf T}|s \rangle~\langle s|z \rangle.
\ea
That is to say:
\ba
\label{repec}
\langle z'|{\bf T}|z\rangle=\left({k \over{\pi\sigma^{2}}}\right)^{1/2}
\left({k \over{2\pi i}}\right)^{1/2} \mbox{e}^{(-i\pi\nu/2)} 
\times \int ds ds' 
\left|{{\partial^{2} l}\over{\partial s \partial s'} }\right|^{1/2}
\exp{\left(ik \Phi(s',s) \right)} 
\ea
where
\ba
\Phi(s',s)=
{i\over 2}z'^{2}+{i\over 2}z^{2}
{i \over {2 \sigma^{2}}}s'^{2}-
{{i\sqrt{2}}\over \sigma}z's'+
{i \over {2 \sigma^{2}}}s^{2}-
{{i\sqrt{2}}\over \sigma}{\bar z}s+l(s',s),
\ea
and $z'=(q'/\sigma-i\sigma p')/\sqrt{2}$ and
$z=(q/\sigma-i\sigma p)/\sqrt{2}$.
The most important contributions come from those points
 $s^{*}$ and $s'^{*}$ that make stationary the phase $\Phi$:
\ba
\label{spa}
{{\partial \Phi}\over{\partial s}}(s^{*},s'^{*})=
{i \over {\sigma^{2}}}s^{*}-
{{i\sqrt{2}}\over \sigma} {\bar z}+
{{\partial l}\over{\partial s}}(s^{*},s'^{*})=0 \nonumber \\
{{\partial \Phi}\over{\partial s'}}(s^{*},s'^{*})=
{i \over {\sigma^{2}}}s'^{*}-
{{i\sqrt{2}}\over \sigma}  z'+
{{\partial l}\over{\partial s'}}(s^{*},s'^{*})=0.
\ea
The solution to (\ref{spa}) satisfying the reality conditions is
\ba
s^{*}=q~~~~~~~{{\partial l}\over{\partial s}}(s^{*},s'^{*})=-p
\nonumber \\
s'^{*}=q'~~~~~~~{{\partial l}\over{\partial s'}}(s^{*},s'^{*})=p'.
\ea
This is a classical trajectory from $(q,p)$ to $(q',p')$. Thus, 
the matrix element $\langle z'|{\bf T}|z\rangle$ will be non zero only if 
 $z$ and $z'$ are connected by the classical dynamics. Let us call 
these points $z_c$ and $z'_c$ and let us calculate the matrix 
element to next order in their neighbourhoods, 
$\langle z_{c}'+\delta z'|{\bf T}|z_{c}+\delta z\rangle$.
To this effect we expand $\Phi(s',s)$ in Eq. (\ref{repec}) to second 
order and, after some algebra:
\ba
\label{phiphi}
\Phi(\delta s',\delta s) \approx 
l(q'_{c},q_{c})
+\left[ -i\delta z' {\bar z'}_{c}-i\delta {\bar z} z_{c}-
{i\over 2} {\bar z}_{c}z_{c}-{i\over 2} {\bar z'}_{c}z'_{c} \right] 
+ \left[
{i\over 4}({\bar z}_{c}^{'2}-z_{c}^{'2}-{\bar z}_{c}^{2}+z_{c}^{2})
\right]
+ \nonumber \\
+\left[
{i\over 2}\delta z'^{2}+{i \over {2 \sigma^{2}}}\delta s'^{2}-
{{i\sqrt{2}}\over\sigma}\delta z'\delta s'+
{i\over 2}\delta {\bar z}^{2}+{i \over {2 \sigma^{2}}}\delta s^{2}-
{{i\sqrt{2}}\over\sigma}\delta {\bar z}\delta s \right. 
\left. +{1\over 2}
\left( {{\partial^{2} l}\over{\partial s'^2}} \delta s'^{2}+
2 {{\partial^{2} l}\over{\partial s' \partial s}} \delta s'\delta s
+{{\partial^{2} l}\over{\partial s^2}}\delta s^{2} \right) 
\right],
\ea
where $\delta s= s-q_c$ and $\delta s'= s'-q'_c$.
We change to new integration variables $\delta s$ and $\delta s'$ and 
obtain:
\ba
\label{repec2}
\langle z'_{c}+\delta z'|{\bf T}|z_{c}+\delta z\rangle \approx
\left({k \over{\pi\sigma^{2}}}\right)^{1/2}
\left({k \over{2\pi i}}\right)^{1/2} \exp{(-i\pi\nu/2)} 
\times 
\int ~d\delta s~ d\delta s'~ 
\left|{{\partial^{2} l}\over{\partial s \partial s'} }\right|^{1/2}
\exp{\left(ik \Phi(\delta s',\delta s) \right)} 
\ea
We now insert Eq. (\ref{phiphi}) in Eq.  (\ref{repec2}). All terms are 
constant with respect to integration except the last one in square 
brackets. The resulting integral is the coherent state representation, 
with  respect to $|\delta z \rangle$, of the linearized map, whose 
generating function is quadratic, which we have introduced in 
Ec. (\ref{reprelin}):
\ba
\label{potprop}
\langle \delta z'|{\bf T}|\delta z \rangle ~=~
{1 \over \sqrt{\bar s_{c}}}~
 \exp \left[ {k \over {2{\bar s}_{c}}}
\left( 
-{\bar r}_{c} \delta z'^{2}+
2 \delta z'  \delta{\bar z}+r_{c} \delta {\bar z}^2
\right)  \right],
\ea
where $r_c$ y $s_c$ are the matrix elements of the linearized map in 
complex coordinates. Finally we arrive at: 
\ba
\label{final}
\langle z'_{c}+\delta z'|{\bf T}|z_{c}+\delta z\rangle \approx
\exp{\left[{-k\over 4}\left(
{\bar z}_{c}^{'2}-z_{c}^{'2}-{\bar z}_{c}^{2}+z_{c}^{2}
\right)\right]} \times \nonumber \\
\times 
\exp{\left[k\left(
\delta z' {\bar z'}_{c}+\delta {\bar z} z_{c}+
{1\over 2} {\bar z}_{c}z_{c}+{1\over 2} {\bar z'}_{c}z'_{c}
\right)\right]} 
\times
\exp{\left(ikl-i{\pi\over 2}\nu\right)}
\langle \delta z'|{\bf T}|\delta z \rangle.
\ea

This result lets us evaluate the matrix element we were looking for, 
$\langle z|{\bf T}^{n}|z\rangle/\langle z|z\rangle$, that will be a sum of 
contributions of periodic points $z_{pp}$ of period $n$ in the 
semiclassical limit:
\ba
{{\langle z|{\bf T}^{n}|z\rangle}\over{\langle z|z\rangle}} \approx
\sum_{pp,n} 
{{\langle z_{pp}+\delta z|{\bf T}^{n}|z_{pp}+\delta z\rangle}\over
{\langle z_{pp}+\delta z|z_{pp}+\delta z\rangle}}.
\ea
To obtain the composition 
$\langle z_{pp}+\delta z|{\bf T}^{n}|z_{pp}+\delta z\rangle$
we use the expression (\ref{final}) and the composition rule of 
Eq. (\ref{compo}). Then
\ba
{{\langle z|{\bf T}^{n}|z\rangle}\over{\langle z|z\rangle}} \approx
\sum_{pp,n}
{1 \over \sqrt{\bar s_{pp}}}~ 
\lambda_{pp} \exp{(ikl_{pp}-i{\pi\over 2}\nu_{pp}) } 
\times
\exp \left[ {k \over {2{\bar s}_{pp}}}
\left( -{\bar r}_{pp} \delta z^{2}+2 \delta z  \delta{\bar z}+
r_{pp} \delta {\bar z}^2 \right) -k \delta z \delta {\bar z} \right],
\ea
where $\lambda_{pp}$ can be calculated by Eq.  (\ref{compo2}), 
$\nu_{pp}=n$ (because of the Dirichlet boundary conditions) and 
$l_{pp}$ is the length of the PO starting from $(q_{pp},p_{pp})$.
As we can see, the matrix element behaves as a gaussian in the vecinities 
of the periodic point. This allows us to write the Husimi representation 
of the $n$-th power of the propagator as a sum of contributions from 
periodic points of period $n$.
Each term of the 
sum is a gaussian packet in phase space whose parameters are related 
to the monodromy matrix in complex coordinates. A PO composed by $n$ 
points will give $n$ {\it different} contributions to this sum, due to 
the fact that the monodromy matrices at each point differ. However, 
the invariant properties of these matrices are the same and the usual 
Gutzwiller-Tabor trace formula \cite{tabor} can be recovered by 
integration.
Of course, a periodic point of period $n$ will contribute also to the 
$rn$ ($r$ natural) powers of the propagator.

We should remark at this point that the different semiclassical 
representations of the propagator in terms od the corresponding 
generating function are only {\it semiclassically} equivalent and
thus can give different results at finite $N$. This is not true for 
the calculation fo the spectral determinant, whose semiclassical 
expression in terms of periodic orbits is the same in all representations.
It is because of this that the different ways of computing eigenfunctions
are not equivalent. For the calculation of $|\phi_{\nu}(s)|^{2}$ the 
closed (but not necesarily periodic) orbits are needed \cite{bogo}. 
For the Wignaer function calculation only periodic points are needed but
each contribution is extended in phase space.
In the present formalism we will obtain the 
Husimi distributions of eigenfunctions in terms of deformed 
localized gaussians 
centered in the periodic points, constructed solely in terms of classical 
information.

\vspace{1.cm}
{\it Symmetries}
\vspace{.4cm}

Our system, the stadium billiard, has two discrete spatial symmetries: 
$R_{x}$ and $R_{y}$, the two reflections with respect of the coordinate 
axes. These spatial symmetries in the domain reflect in the border and, 
thus, in the classical and quantum map on it. Their action on the Birkhoff
coordinates of phase space $(q,p)$ is 
\ba
\label{simets}
R_{x} (q,p) ~\rightarrow ~ (L-q,-p),~~~ 
R_{y} (q,p) ~\rightarrow ~ \left({L \over 2}-q,-p\right),~~~ 
R_{x} R_{y} (q,p) ~\rightarrow ~ \left({L\over 2}+q,+p\right).
\ea
In order to have coherent states on the border with correct symmetries we 
need to project them using ${\bf R}_x$ and ${\bf R}_y$, the unitary 
representations of the symmetries ${\bf R}_{x}|x,y \rangle=|-x,y\rangle$ 
and ${\bf R}_{y}|x,y \rangle=|x,-y\rangle$. Then we define
\ba
|z_{\sigma_{x} \sigma_{y}}\rangle~=~
\left( {{1+\sigma_{x} {\bf R}_{x}}\over 2} \right)
\left( {{1+\sigma_{y} {\bf R}_{y}}\over 2} \right) 
{{|z\rangle}\over\sqrt{\langle z|z\rangle}},
\ea
where $\sigma_{x}, \sigma_{y}=\pm 1$ and ${\bf R}_x$ and ${\bf R}_y$ 
move the center of the coherent state according to Ec. (\ref{simets}).

In this way, the diagonal matrix elements of the propagator in symmetrized 
coherent state representation are
\ba
\label{cssim}
{1\over{\langle z|z\rangle}}
\langle z|{\bf T} \left( {{1+\sigma_{x} {\bf R}_{x}}\over 2} \right)
\left( {{1+\sigma_{y} {\bf R}_{y}}\over 2} \right) |z\rangle 
={1\over{4\langle z|z\rangle}} ( 
\langle z|{\bf T}|z\rangle+
\sigma_{x} \langle z|{\bf T}{\bf R}_{x}|z\rangle+
\sigma_{y} \langle z|{\bf T}{\bf R}_{y}|z\rangle
+\sigma_{x} \sigma_{y} \langle z|{\bf T}{\bf R}_{x}{\bf R}_{y}|z\rangle ).
\ea

We have already calculated $\langle z|{\bf T}^{n}|z\rangle$.
We still have to calculate the other three contributions,  
$\langle z|{\bf T}^{n}{\bf R}_{x}|z\rangle$, $\langle z|{\bf T}^{n}{\bf R}_{y}|z\rangle$ and 
$\langle z|{\bf T}^{n}{\bf R}_{x}{\bf R}_{y}|z\rangle$.
We can conclude using the results we have already obtained that each of 
them will be a sum of gaussians centered in those points $z$ that the 
dynamics connects with their symmetric partners, 
$R_{x}z$, $R_{y}z$, $R_{x}R_{y}z$, respectively.
These points belong to POs whose periods are $2n$ which are symmetric 
under the operations 
$R_{x}$, $R_{y}$, $R_{x}R_{y}$, respectively.
The increment $R \delta z$ with respect to $R z$ 
($R \equiv R_{x}, R_{y}, R_{x}R_{y}$) is related to the increment 
$\delta z$ with respect to $z$ through: 
\ba
R \delta z=t_{R} \delta z,~~~~~~~
t_{R}=\left\{ \begin{array}{lc} -1 \mbox{ if }R=R_{x} \\
-1 \mbox{ if }R=R_{y} \\ 1 \mbox{ if }R=R_{x}R_{y}. \\
\end{array} \right.
\ea
Thus we arrive at:
\ba
{{\langle z|{\bf T}^{n}{\bf R}|z\rangle}\over{\langle z|z\rangle}} \approx
\sum_{pp,2n}
{1 \over \sqrt{\bar s_{pp}}}~ 
\lambda_{pp} \exp{(ikl_{pp}-i{\pi\over 2}\nu_{pp})}
\times
\exp \left[ {k \over {2{\bar s}_{pp}}}
\left( -{\bar r}_{pp} \delta z^{2}+2 t_{R} \delta z  \delta{\bar z}+
r_{pp} \delta {\bar z}^2 \right) -k \delta z \delta {\bar z} \right],
\ea
where the sum goes over the periodic points of period $2n$ that belong 
to POs symmetric under $R$.
The quantities $s_{pp}$, $r_{pp}$, $l_{pp}$, $\nu_{pp}$ and $\lambda_{pp}$ 
are calculated along the trajectory that connect $z$ to $Rz$, i.e., half 
PO.

\section{Semiclassical eigenfunctions for the stadium}
\label{secauto}

We use the semiclassical approach we introduced above for the stadium 
billiard. We choose odd-odd symmetries ($\sigma_{x}=\sigma_{y}=-1$) and 
$\sigma=2$. We have periodic points with desymmetrized period up to 8 
(around 800). We have used the symbolic dynamics developed by Biham and 
Kvale \cite{biham} to obtain them. 
The wave number $k$ is related to the maximum period used 
in the expansion (\ref{husimm}) by $P(k)={L\over{2\pi}} k \approx 0.4~ k$.
In this way we can obtain semiclassical approximations of eigenfunctions 
of wave number $ k \leq 20$. 

In Fig. \ref{propquan} we show  the phase space representations of the 
first six powers of Bogomolny's ${\bf T}$ operator for $k=20$; in Fig. \ref{propsemi}, the 
semiclassical approximations. We see that the exact representations show 
global maxima in the bouncing ball region that cannot be reproduced 
semiclassically for the lowest powers. However, the overall 
semiclassical behaviour is very close to that of the exact 
representations. (Because of the symmetries we chose, we have no 
semiclassical approximation to the first power of the operator
because the contribution of the only periodic point of period 1 is zero.)

We select two energy ranges: $k \in [19.1,20.0]$ and 
$k \in [20.5,21.3]$. There are 4 eigenenergies in each of these ranges.
We show the absolute value of the secular determinant, $|P(k)|$, for 
each of them in Figs. \ref{detsecu1} and \ref{detsecu2}. The full line 
is the semiclassical approximation, the dashed line is the secular 
determinant for Bogomolny's operator. The vertical lines are the exact 
quantum $k$ eigenvalues calculated by the scaling method \cite{vergini}.
We see a good approximation when we use the periodic point expansion.
The agreement shows that in this region the spectrum is well 
represented semiclassically.
(The discontinuities come from the change in dimension of the operator).

To keep the method consistent we evaluated the semiclassical Husimi 
expansion in those values of $k$ that minimize the 
semiclassical secular determinant. 
We see in Figs. \ref{semih1} and \ref{semih2} the exact eigenfunctions 
(first column) and their corresponding semiclassical Husimi representations
(second column) obtained as the real part of Eq. (\ref{husimm}). 
The global behaviour is well reproduced; however, the finer details are 
hard to mimic. The bouncing ball region is problematic: in some 
functions (e.g., $k=21.16$) some probability leaks to this region. 
Probably POs with longer periods that approximate the bouncing ball
orbits could make a better picture for this region.

One of the advantages of formula (\ref{proyec}) for the projector is that 
it has no singularities between eigenvalues. It is possible to study 
continualy its behaviour as a function of $k$ in order to see its 
sensitivity to changes in $k$. Some properties of the exact distribution 
as a function of $k$ are:
\begin{itemize}
\item The distribution is positive at the eigenvalues $k_n$.
\item The distribution has $N(k)$ zeros at eigenvalues $k_n$.
\end{itemize}
These two properties follow from the fact that 
${\cal H}_{\psi_{\nu}}(z,{\bar z})$ is the modulus of an analytic 
function.
The distributions between eigenvalues can become negative. In 
particular, it can be shown that at the value of $k$ that maximizes 
$P(k)$, the distribution is constant.
This properties can be used to control the semiclassical approximations.

In Figs. \ref{movie1} and \ref{movie2} we show the behaviour of the 
distribution $\langle z|{\bf g}(k)|z\rangle/\langle z|z\rangle$ between 
the 
semiclassical eigenvalues $k=19.18$ and $k=19.38$. When $k=19.18$ the 
distribution is positive and has well defined minima that approach zero.
As we move away from the eigenvalue, the distribution changes smoothly. 
Initialy it moves away from the plane $g=0$ in the positive direction, 
then it comes back and turns negative. During this ``evolution'' it 
flattens visibly and we can't discern its features. At $k=19.32$, 
aproximately the maximum of the secular determinant, see Fig. 
\ref{detsecu1}, the distribution is constant. At the semiclassical 
eigenvalue $k=19.38$ the distribution is positive again with well 
defined minima.  

We can see from Figs. \ref{semih1} and \ref{semih2} that the 
semiclassical approximation is relatively good. It is not trivial to 
obtain a positive defined distribution with $N$ zeroes adding several 
hundreds of gaussians, each with its phase and deformation.

\section{Conclusions}
\label{secconclu}

Using Fredholm theory we have given a very compact and representation 
independent derivation of the projector on a single eigenfunction for
unitary quantum maps. Expressing the projector in the coherent state 
basis we wrote a semiclassical expression for the Husimi 
distributions of the billiard's eigenfunctions. Each periodic point 
contributes with a gaussian centered in it whose parameters are calculated 
only with classical information.  
We should not underestimate the difficulties and complexities 
inherent to this method. Hundreds of gaussian contributions have to 
conspire to make a positive definite distribution with $N$ that approximate 
the quantum Husimi distributions.

The projector (\ref{proyec}) can be represented in coordinate space. 
We obtain $\langle q| \psi \rangle\langle\psi|q\rangle$, whose 
semiclassical approximation can be directly compared to the probability 
density in the section. This representation has an additional difficulty, 
since the semiclassical approximation is written as a sum over closed 
trajectories, periodic or not, in configuration space. Those that are not 
periodic are more in number and more difficult to find. Anyway, we can apply 
our scheme for Bogomolny's ${\bf T}(q',q)$ operator and compare the results with 
the exact quantum calculation. In Fig. \ref{psi2bogo} we see that the 
approximation is excelent at this level.

The maximum period $P$ in the expansions is related to the energy in the 
way  $P \approx 0.4 k$. Due to the exponential proliferation of orbits in
chaotic systems, the method cannot be applied for arbitrarily high energies.
The measure of this proliferation is the topologic entropy $W$ which 
relates the number $N_{P}$ of POs of a given period $P$ with the period 
itself, $N_{P}=\exp(WP)$ \cite{liberman}. For the stadium, $W \approx 0.94$.
Then, for $k \approx 100$ we need POs of periods up to $P=40$, whose number
is  $N_{P} \approx \exp(0.94 \times 40) \approx 10^{17}$!

The Fredholm method we developed is a first step and shows that the 
eigenfunctions can be described as expansions in terms of the periodic 
points of the underlying classical system. It eliminates the divergencies 
associated that the schemes based on smoothings in energy have. 
However, the exponential divergence of periodic orbits poses a serious 
practical problem, as discussed in the previous paragraph. 
This method can only become practical for large $k_{\nu}$ if some way of 
selecting a few ``important'' orbits at each value of $k$ can be developed.
Some results in this direction have been obtained by Vergini and Carlo
\cite{carlo,vergini}.

\appendix
\section{Complex phase space}
\label{seccomplex}

We introduce the following symplectic transformation $Z$, depending upon 
parameter $\sigma$, acting on a point of classical phase space $(q,p)$
\cite{bargmann}:
\ba
\label{defz}
\left( \begin{array}{lc} z \\ p_{z} \\ \end{array} \right) ~=~
Z \left( \begin{array}{lc} q \\ p \\ \end{array} \right) ~=~
\left( \begin{array}{lclc} 1/\sqrt{2} \sigma & -i\sigma/\sqrt{2} \\
-i/\sqrt{2}\sigma & \sigma/\sqrt{2} \\ \end{array} \right)
\left( \begin{array}{lc} q \\ p \\ \end{array} \right).
\ea
Imposing reality conditions on the inverse transformation we see that 
${\bar z}=i p_{z}$. A linear transformation 
$M=\left( \begin{array}{lclc} a & b \\ c & d \\ \end{array} \right)$ 
in $(q,p)$ phase space 
has a representation $M_{z}$ in $(z,p_{z})$ phase space by conjugation 
with $Z$
\ba
M_{z}=Z M Z^{-1}=
\left( \begin{array}{lclc} {\bar s} & -ir \\ 
i{\bar r} & s \\ \end{array} \right)
~~~~~~~\mbox{with}~~\left\{ 
\begin{array}{lc} 
s~=~{1 \over 2} \left[(a+d)-i\left({b \over {\sigma^{2}}}-\sigma^{2} c
\right)\right] \\
r~=~{1 \over 2} \left[(d-a)+i\left({b \over {\sigma^{2}}}+\sigma^{2} c
\right)\right] 
\end{array} \right. .
\ea

This $(z,p_{z})$ phase space allows a passage to quantum mechanics. 
This is done in a Hilbert Bargmann space by introducing operators 
$\bf z$ and $\bf p_{z}$ that satisfy the conmutator relations
\ba
[{\bf z},{\bf p_{z}}]={i \hbar} ,~~~~[{\bf z},{\bf z}]=
[{\bf p_{z}},{\bf p_{z}}]=0.
\ea
Any vector $|\psi \rangle$ in Hilbert space can be represented in this new 
space as $\langle z|\psi \rangle=
\int dq \langle z|q\rangle \langle q| \psi \rangle$, where the coherent 
states are
$\langle z | q \rangle = (1/ (\pi \hbar \sigma^{2}))^{1/4}
~ \exp{ \left( -(1/\hbar) \left(
z^{2}/2 + q^2 /(2\sigma^2) - \sqrt{2} z q / \sigma \right) \right) }$.
The scalar product is
$\langle \psi_{1}|\psi_{2} \rangle=
\int ~{\bar \psi}_{1}(z) ~\psi_{2}(z)~d\mu (z)$ with norm 
$d\mu (z)~=~{1 \over \pi} \exp \left(- z{\bar z /\hbar} \right)
d\mbox{Re}(z)~d\mbox{Im}(z)$.

We now define the Husimi representation of a vector $\psi$ as
\ba
{\cal H}_{\psi}(z)~\equiv~
{{| \langle z | \psi \rangle  |^{2}}\over{\langle z | z \rangle}}.
\ea
It is a real positive function for every $z$ in the complex plane.

The representation of a linear symplectic transformation $M$ in phase space 
in terms of a unitary operator of Hilbert Bargmann space is \cite{moshinsky}
\ba
\label{reprelin}
M~=~
\left( \begin{array}{lclc} a & b \\ c & d \\ \end{array} \right)
~~~~~~\rightarrow
~~~~~ \langle z'|{\bf U}(M)| z \rangle
={1 \over \sqrt{|\bar s}|}~\exp \left(  {-i \over 2} 
\mbox{arg}({\bar s})
\right)
\exp \left[ {k \over {2{\bar s}}}
\left( -{\bar r} z'^{2}+2 z' {\bar z}+r {\bar z}^2 \right) 
\right].
\ea
This representation is up to a phase \cite{moshinsky,little} and its 
composition law is 
\ba
\label{compo}
\int \langle z'|{\bf U}(M_{1})| z \rangle
\langle z|{\bf U}(M_{2})| z'' \rangle ~ d\mu (z)~
=~\lambda(M_{1},M_{2},M_{1}M_{2}) ~
\langle z'|{\bf U}(M_{1}M_{2})| z'' \rangle,
\ea
with
\ba
\label{compo2}
 \lambda(M_{1},M_{2},M_{1}M_{2}) =
\exp \left[ {i \over 2} \left(
\mbox{arg}({\bar s})-\mbox{arg}({\bar s}_{1})-\mbox{arg}({\bar s}_{2}) 
-\mbox{arg}\left( {{\bar s} \over
{{\bar s}_{1} {\bar s}_{2}  }}\right) 
\right)\right]~=~\pm 1.
\ea
The accumulated phase due to succesive transformations leads to the 
Maslov index of the trajectory.

In case the phase space shows periodicity in coordinate or momentum, we 
have to periodize the coherent states as in \cite{voros}.



\newpage
\section*{Figure captions}

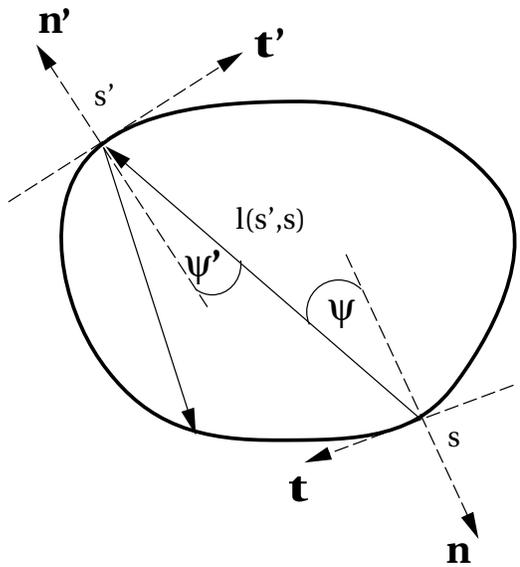
\begin{figure}[h!]
\caption[]{\label{figbillar} Geometry of the billiard.}
\end{figure}
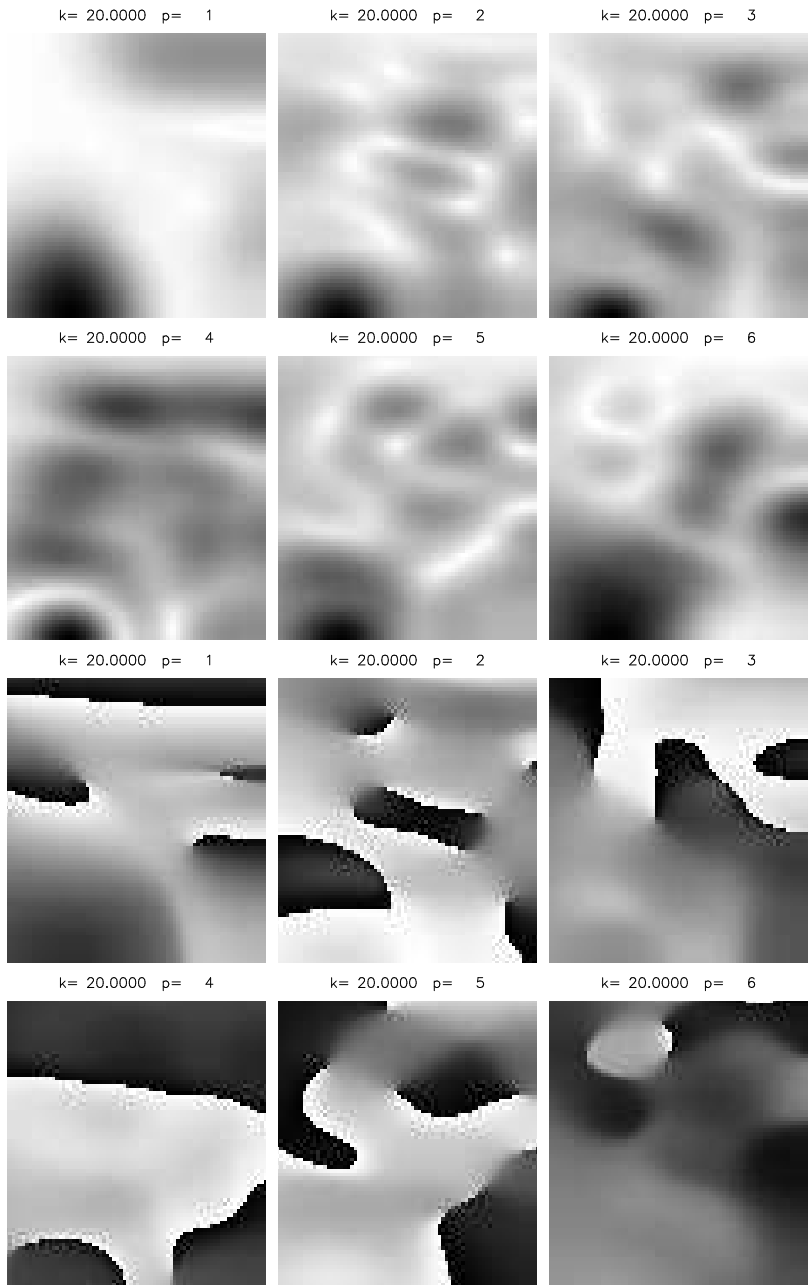
\begin{figure}[h!]
\caption[]{\label{propquan} Phase space representations for the first 
six powers of Bogomolny's operator. For each power we show modulus 
(rows 1 and 2) and phase (rows 3 and 4).}
\end{figure}
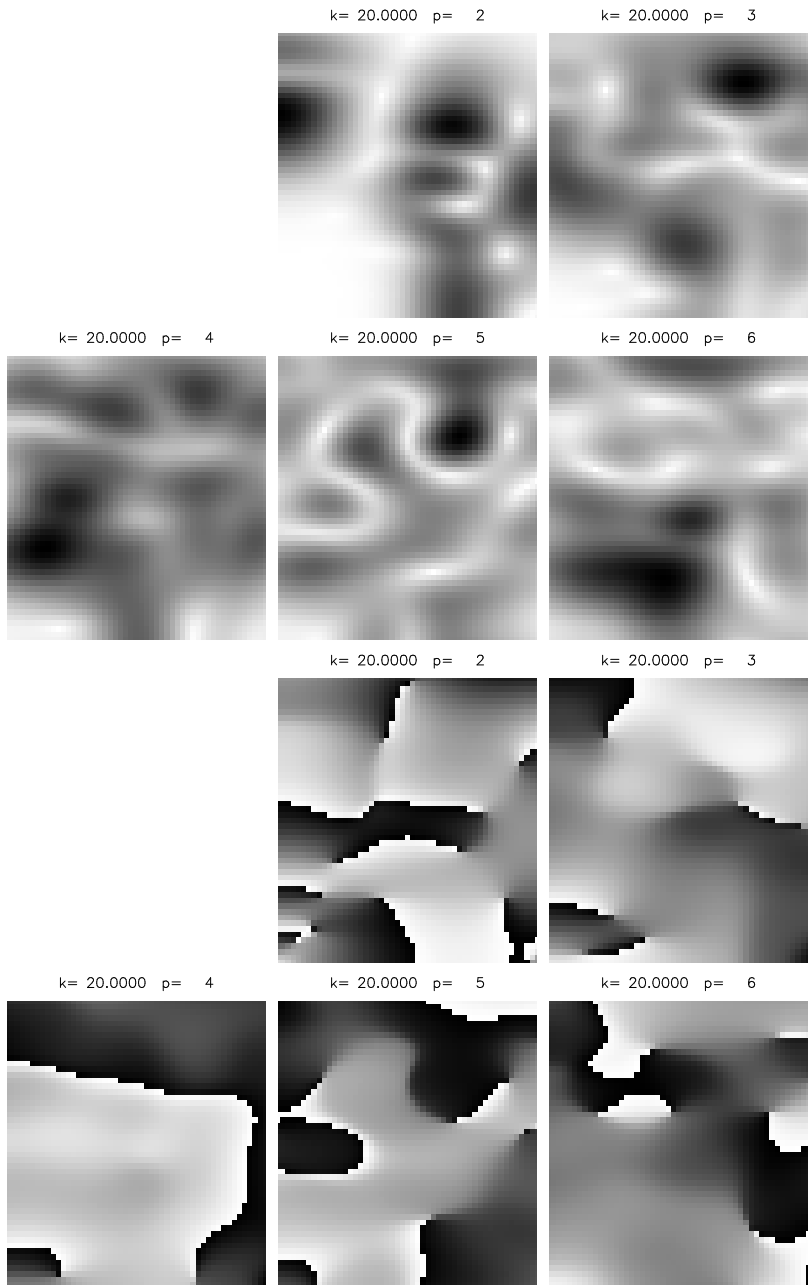
\begin{figure}[h!]
\caption[]{\label{propsemi} Semiclassical
phase space representations for the first 
six powers of Bogomolny's operator. For each power we show modulus 
(rows 1 and 2) and phase (rows 3 and 4).}
\end{figure}
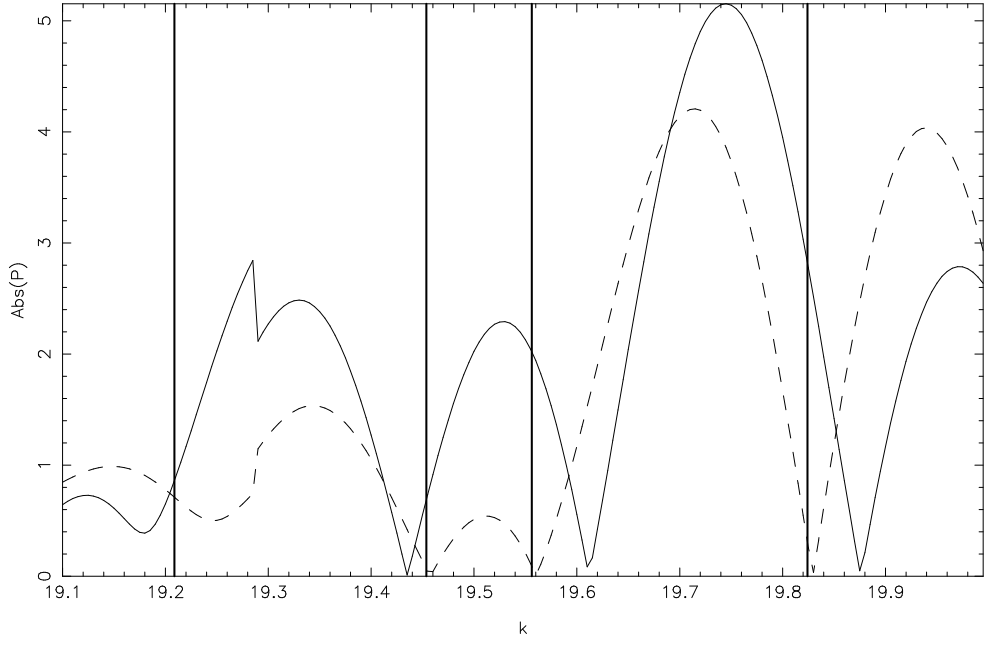
\begin{figure}[h!] 
\caption[h]{\label{detsecu1} Secular determinant. In full line we show 
the PO's approximation; in dashed line, the exact by using Bogomolny's
operator. For the semiclassical approximation we summed up to period 8.}
\end{figure}
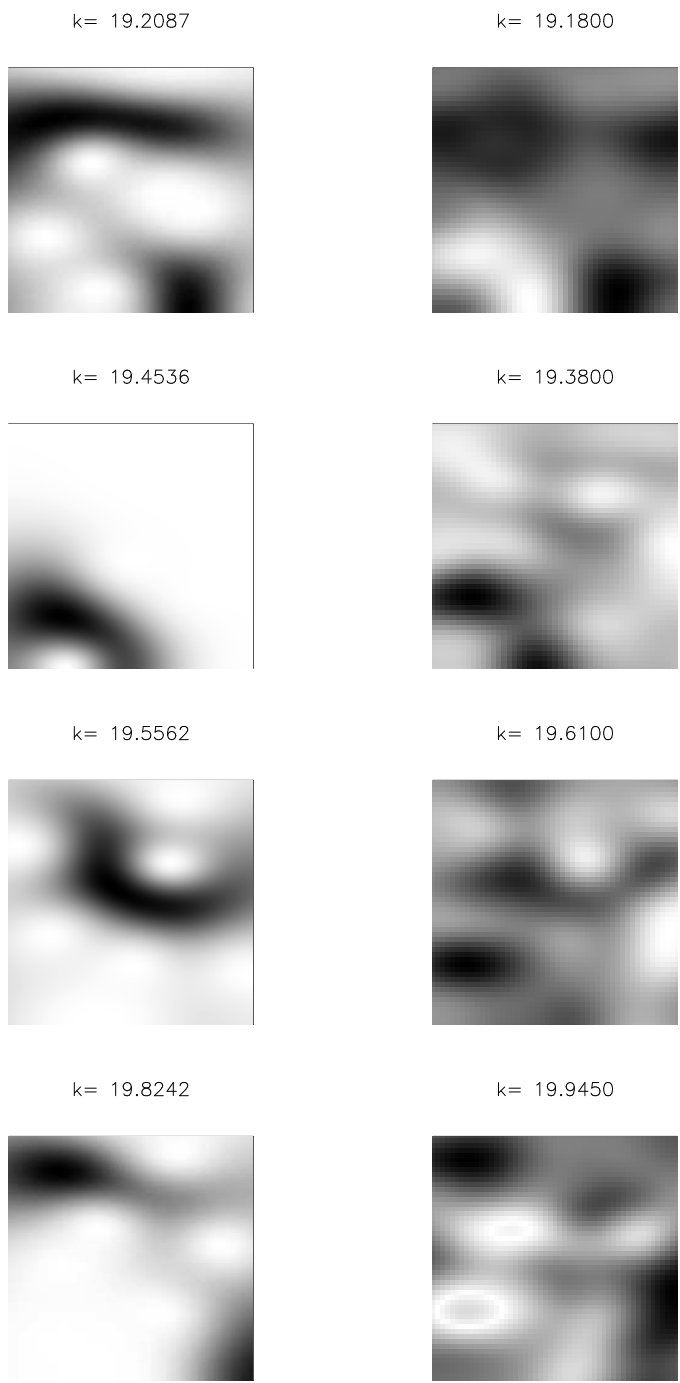
\begin{figure}[h!] 
\caption[h]{\label{semih1} Husimi representations of stadium eigenfunctions
(left panels) and their semiclassical approximations (right panels) 
for the energy range of Fig. \ref{detsecu1}.
For the semiclassical approximation we summed up to period 8.}
\end{figure}
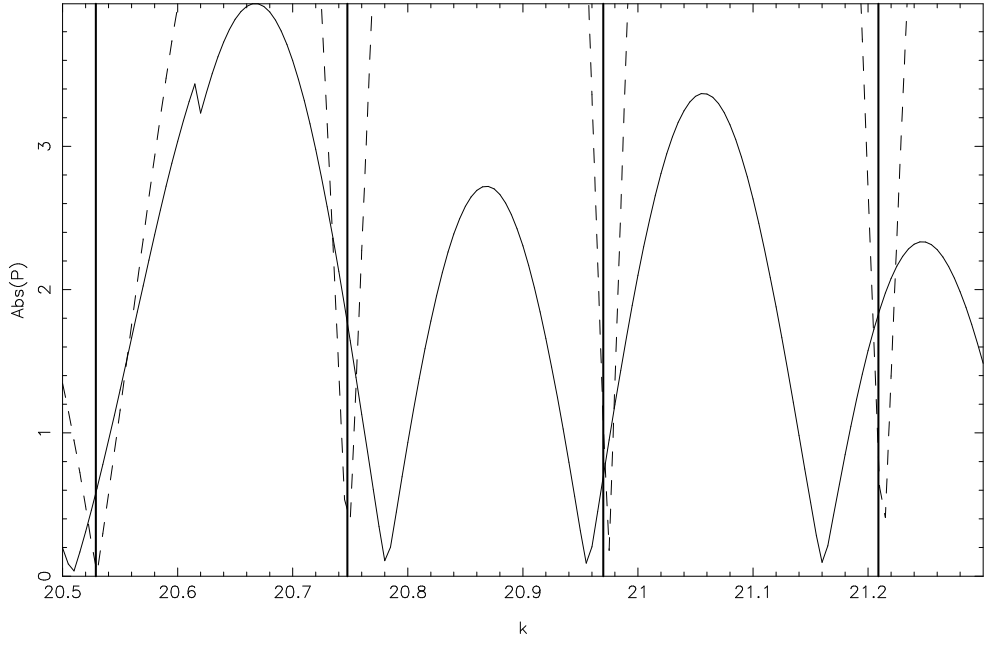
\begin{figure}[h!] 
\caption[h]{\label{detsecu2}Secular determinant. In full line we show 
the PO's approximation; in dashed line, the exact by using Bogomolny's
operator. For the semiclassical approximation we summed up to period 8.}
\end{figure}
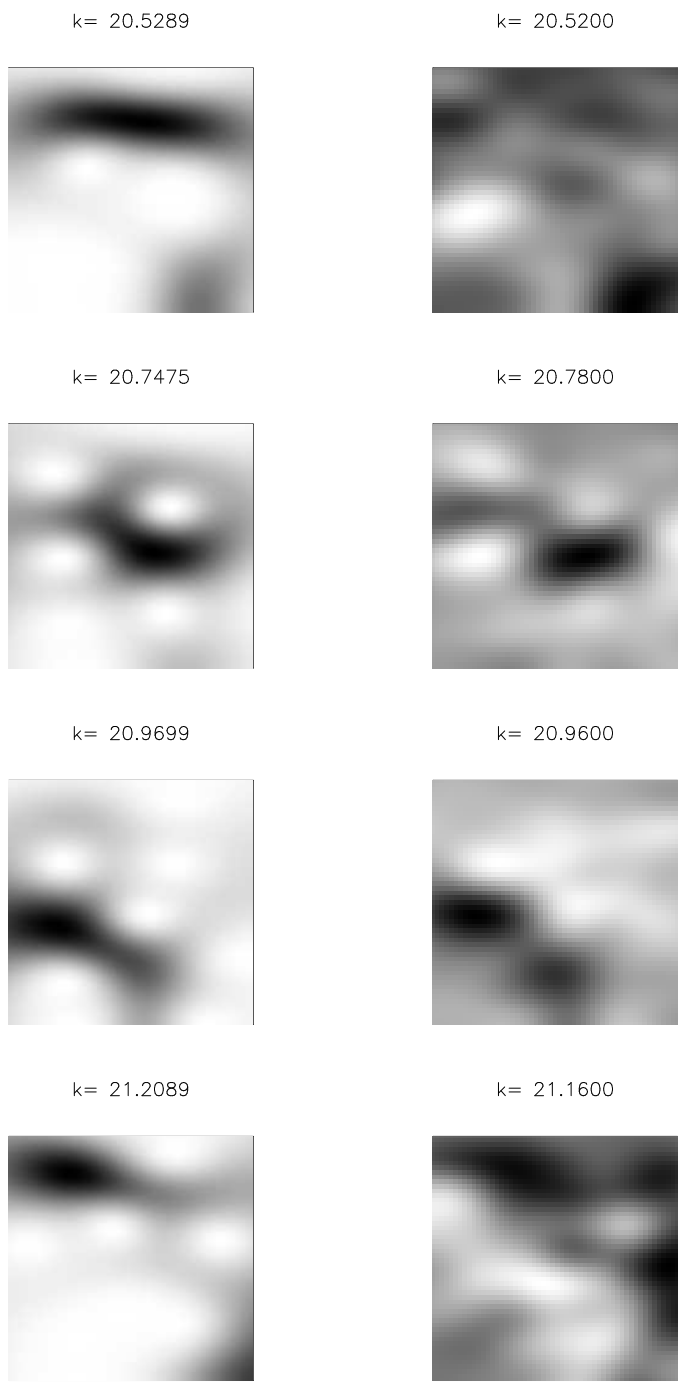
\begin{figure}[h!] 
\caption[h]{\label{semih2}Husimi representations of stadium eigenfunctions
(left panels) and their semiclassical approximations (right panels) 
for the energy range of Fig. \ref{detsecu2}.
For the semiclassical approximation we summed up to period 8.}
\end{figure}
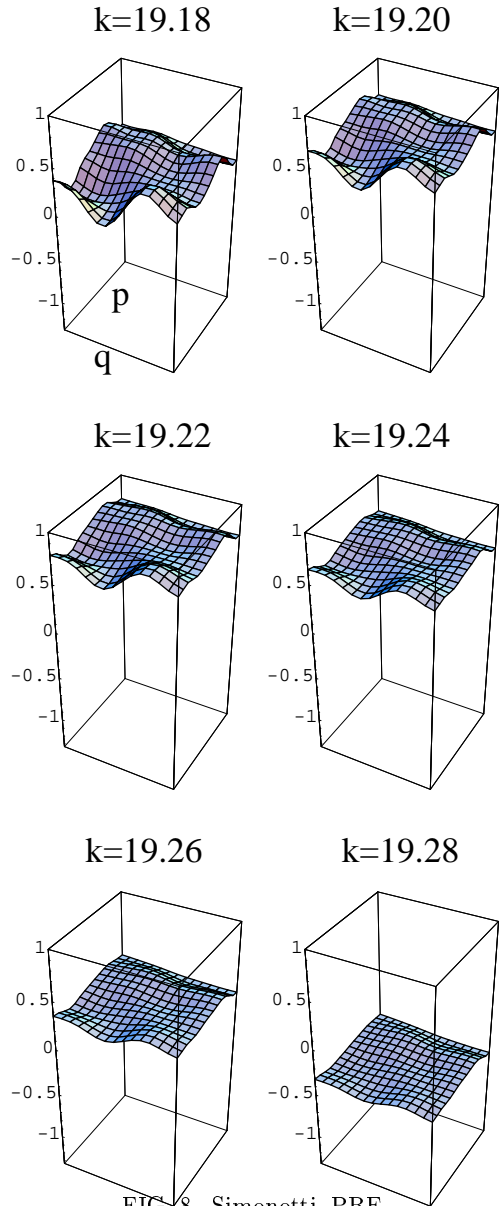
\begin{figure}[h!] 
\caption[h]{\label{movie1} Variation of the normalized Green function in 
coherent state representation between $k=19.18$ and $k=19.28$. $k=19.18$
is a semiclassical eigenvalue.}
\end{figure}
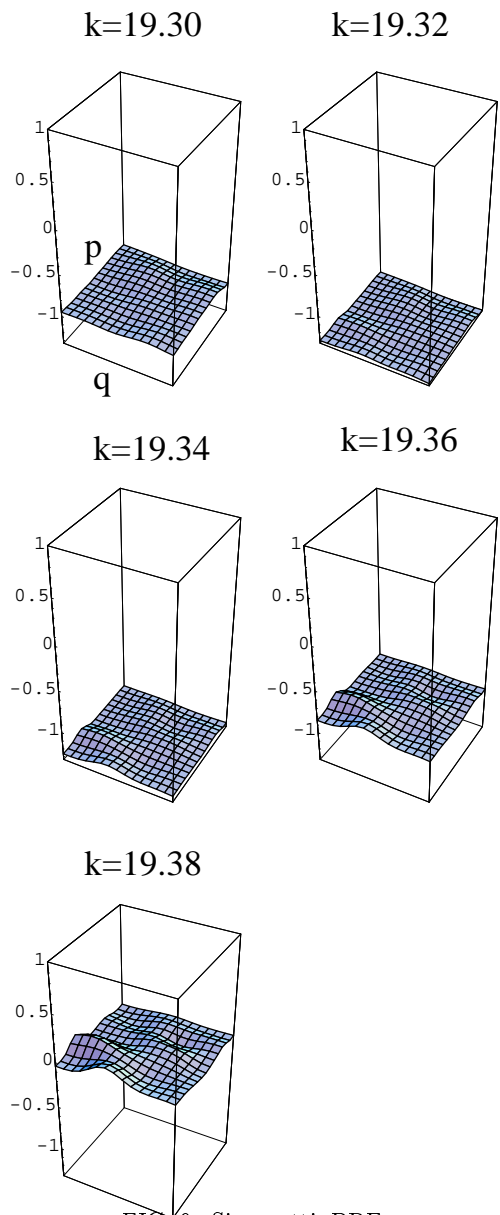
\begin{figure}[h!] 
\caption[h]{\label{movie2}
Continuation of Fig. \ref{movie1}.
Variation of the normalized Green function in 
coherent state representation between $k=19.28$ and $k=19.38$. $k=19.38$
is a semiclassical eigenvalue.}
\end{figure}
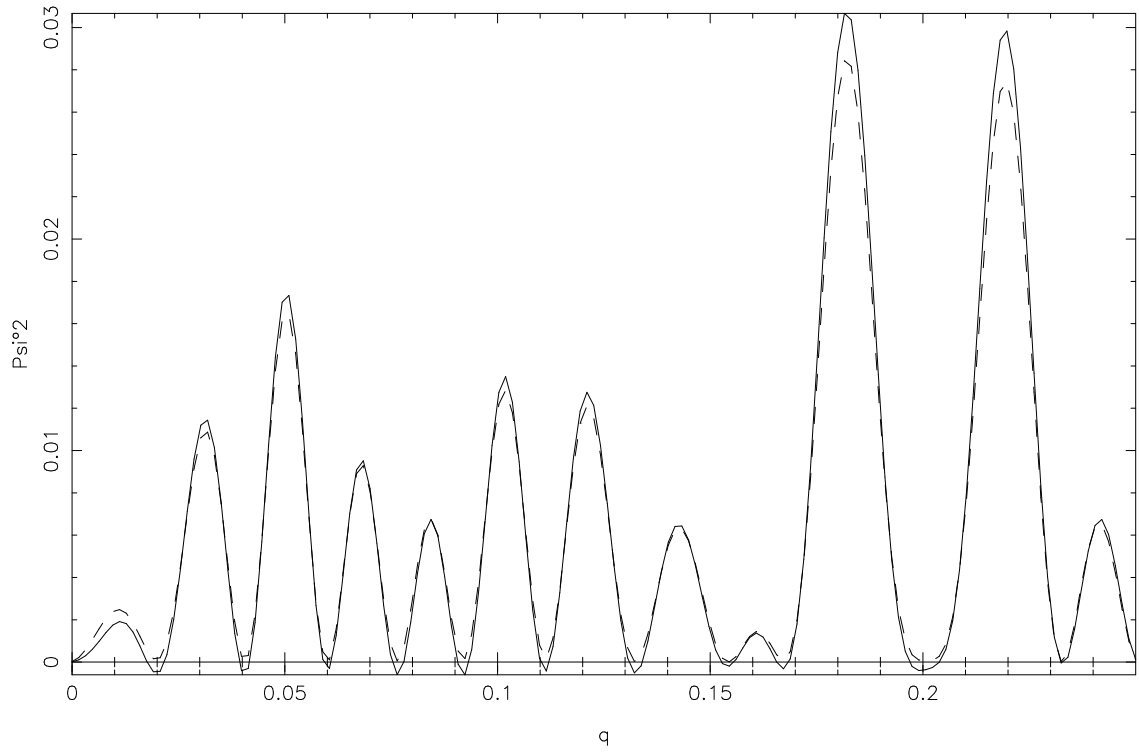
\begin{figure}[h!] 
\caption[h]{\label{psi2bogo} Exact eigenfunction (dashed line) for 
$k=20.5289$ and its approximation by using Bogomolny's operator in 
the normalized Green function (full line).}
\end{figure}



\newpage
\begin{thebibliography}{99}

\bibitem{bogo} E. B. Bogomolny, Nonlinearity {\bf 5}, 805 (1992).

\bibitem{vorosspec} A. Voros, J. Phys. A {\bf 21}, 685 (1988).

\bibitem{keating} J. P. Keating and M. Sieber, Proc. R. Soc. Lond. 
A {\bf 447}, 413 (1994).

\bibitem{heller} E. J. Heller, in {\it Chaos and Quantum Physics}, 
Proceedings from Les Houches 1989 (North-Holland, Amsterdam, 1989).

\bibitem{bogoscars} E. B. Bogomolny, Physica D {\bf 31}, 169 (1988). 

\bibitem{berryscars} M. V. Berry, Proc. Roy. Soc. A {\bf 423}, 
219 (1989).

\bibitem{agam} O. Agam and S. Fishman, Phys. Rev. Lett. {\bf 73}, 
806 (1994).

\bibitem{smilansky} D. Klakow and U. Smilansky, J. Phys. A {\bf 29}, 
3213 (1996).

\bibitem{simonotti} F. P. Simonotti, E. Vergini and M. Saraceno, 
Phys. Rev. E {\bf 56}, 3859 (1997).

\bibitem{kaplan} L. Kaplan and E. J. Heller, Ann. Phys. (NY) {\bf 264}, 
171 (1998).

\bibitem{berryconj} M. V. Berry, J. Phys. A {\bf 10}, 2083 (1977).

\bibitem{vorosconj} A. Voros, in {\it Stochastic Behaviour in 
Classical and Quantum Hamiltonian Systems}, eds. G. Casati y G. Ford,
Lectures Notes in Physics {\bf 93} (Springer, berlin, 1979), p. 326. 

\bibitem{shnirelman} A. Shnirelman, Usp. Mat. Nauk. {\bf 29}, 
181 (1974).

\bibitem{fishman} S. Fishman, B. Georgeot and R. E. Prange, J. Phys. A 
{\bf 29}, 919 (1996).

\bibitem{smithies} F. Smithies, {\it Integral equations}, Cambridge 
Tracts in Mathematics and Mathematical Physics {\bf 49}, (Cambridge 
University Press, Cambridge, 1962).

\bibitem{ozorio} A. Ozorio de Almeida, Phys. Rep. {\bf 295}, 267 (1998).

\bibitem{liberman} A. Lichtemberg and M. Lieberman, {\it Regular and
Stochastic motion}, Springer - Verlag, (1983).

\bibitem{voros} J. M. Tualle and A. Voros, Chaos Solitons  
Fractals, {\bf 5}, 1085 (1995).

\bibitem{smilden} U. Smilansky, in {\it Mesoscopic Quantum Physics}, 
Proceedings from Les Houches 1994 (North Holland, Amsterdam, 1995).

\bibitem{gutz} Gutwiller, J. Math. Phys. {\bf 12}, 343 (1971)

\bibitem{bargmann} V. Bargmann, {\it Group representations on 
Hilbert spaces of analytic functions} in ``Analytic Methods in 
Mathematical Physics" (R. P. Gilbert and R. G. Newton eds.), 
Gordon and Breach, New York, 1968.

\bibitem{biham} O. Biham and M. Kvale, Phys. Rev. A {\bf 46}, 
6334 (1992).

\bibitem{moshinsky} P. Kramer, M. Moshinsky and T. H. Seligman, 
{\it Complex Extensions of Canonical Transformations and Quantum 
Mechanics}, in ``Group Theory and its Applications", Academic Press
(1975).

\bibitem{little} R. G. Littlejohn, Phys. Rep. {\bf 13}, 193 (1986).

\bibitem{vergini} E. Vergini and M. Saraceno, Phys. Rev. E {\bf 52}, 
2204 (1995).


\bibitem{boasman} P. A. Boasman, Nonlinearity {\bf 7}, 485 (1994).

\bibitem{berry} M. V. Berry and M. Wilkinson, 
Proc. Roy. Soc. London, Ser. A {\bf 392}, 15 (1984).

\bibitem{tanner} G. Tanner, J. Phys. A {\bf 30}, 2863 (1997).

\bibitem{tabor} M. Tabor, Physica D {\bf 6}, 195 (1982).


\bibitem{monastra} A. Monastra and M. Saraceno, unpublished.

\bibitem{vergini2} E. Vergini, submitted to J. Phys. A.

\bibitem{carlo} G. Carlo and E. Vergini, submitted to J. Phys. A.

\end{thebibliography}
\end{document}